\documentclass{appolb}
\usepackage{epsfig}
\begin{document}

\title{FEATURES OF EXCEPTIONAL POINTS AND THE CONTINUUM SPECTROSCOPY
\thanks{Presented at the Zakopane Conference on Nuclear Physics,
September 1-7, 2008}}
\author{J. Oko{\l}owicz
\address{Institute of Nuclear Physics, Radzikowskiego 152, PL-31342 Krak\'ow, Poland}
\and
M. P{\l}oszajczak
\address{Grand Acc\'{e}l\'{e}rateur National d'Ions Lourds (GANIL), CEA/DSM -- CNRS/IN2P3, BP 5027, F-14076 Caen Cedex 05, France}
}
\maketitle

\begin{abstract}
We discuss observable features of exceptional  points and the resonance spectroscopy of 
$^{16}$Ne using the Shell Model Embedded in the Continuum.
\end{abstract}


\section{Introduction}

Small quantum systems embedded in the continuum of scattering and decay channels, are intensely studied in various fields of Physics (nuclear physics, atomic and molecular physics, nanoscience, quantum optics, etc.). These  different open quantum systems (OQS), in spite of their specific features, exhibit certain generic properties such as the segregation of time scales \cite{Kle85,Dro00,[Oko03]}, the alignment of near-threshold states with decay channels \cite{Ike68,Cha06,Dob07}, the instability of SM eigenstates at the channel threshold \cite{Mic07,Oko08}, or the resonance crossings \cite{Rot00,[Oko03]}. All those counterintuitive properties result from the non-hermitian nature of an eigenvalue problem in OQSs.

Resonances are commonly found in various quantum systems.  A special feature of nuclear OQS problem is the strong configuration mixing which has to be handled in both resonant states and non-resonant scattering continuum. A suitable framework is provided by the continuum shell model  (CSM) \cite{[Oko03]}, either the real-energy CSM \cite{SMEC,[Oko03],SMEC_2p} or the  complex-energy CSM \cite{Bet02,Mic02}. 

One of those salient features of OQSs is related to the eigenvalue degeneracies. In the past, level degeneracies have been much studied in connection with avoided crossings in spectra, focusing mainly on  the topological structure of Hilbert space and associated geometrical phases \cite{Ber84,Lau94}. Among these degeneracies, one finds an exceptional point (EP) \cite{Kat95,Zir83,Hei91}, where two Riemann sheets of eigenvalues  are entangled by the square-root type of singularity. 

In many-body systems, EPs have been studied only in schematic models, such as the Lipkin model \cite{Hei91a},   the interacting boson model  \cite{Cej06}, and the 3-level pairing model \cite{Duk09} from the general class of Richardson-Gaudin models  \cite{Duk04, Ort05}. In this paper, we shall present first studies of an EP and its consequences in a realistic nuclear CSM. 

\section{Shell Model Embedded in the Continuum}

Shell Model Embedded in the Continuum (SMEC) is a recent realization of the 
real-energy CSM \cite{SMEC,SMEC_2p}. The total function space in this model consists of 
two sets: the set of square-integrable functions used in the standard nuclear Shell Model (SM) and the set of scattering states into which the SM states are embedded. These two sets are defined by solving the Schr\"odinger equation, separately for discrete SM states (the CQS):
 $H_{SM}\Phi_i=E_i^{(SM)}\Phi_i$ and for scattering states (the environment): 
$ \sum_{c'}(E-H_{cc'})\zeta_E^{c'(+)}=0$. Here, $H_{SM}$ is the SM Hamiltonian and 
$H_{cc}=H_0+V_c$ is the standard Hamiltonian used in coupled-channel (CC) calculations. Channels are determined by the motion of unbound particle in a state $l_j$ relative to the  $A-1$ nucleus with all nucleons on bounded single-particle (s.p.) orbits in a CQS eigenstate (a SM state) $\Phi_j^{A-1}$.  

By means of two functions sets : ${\cal Q}\equiv \{\Phi_i^{A}\}$,
${\cal P}\equiv \{\zeta_E^{c(+)}\}$, one can define the corresponding
projection operators ${\hat Q}$, ${\hat P}$, and the projected Hamiltonians : 
${\hat Q}H{\hat Q}\equiv H_{QQ}$, ${\hat P}H{\hat P}\equiv H_{PP}$.  
Assuming ${\cal Q}+{\cal  P}=I_d$, one can determine the third function set by solving the CC equations with the source term: $  \omega_i^{(+)}(E)=G_P^{(+)}(E)H_{PQ}\cdot \Phi_i~,$
where $G_P^{(+)}(E)={\hat P}(E-H_{PP})^{-1}{\hat P}$ is the Green's function for the motion of a single nucleon in the ${\cal P}$ subspace, $E$ is the total energy of the nucleus $A$, and 
$H_{PQ}\equiv {\hat  P}H{\hat Q}$. $\omega_i^{(+)}$ is the continuation of a SM wave function 
$\Phi_i^A$ into the continuum ${\cal P}$.  

Using the three function sets : $\{\Phi_i^A\}$, $\{\zeta_E^{c(+)}\}$ and $\{\omega_i^{(+)}\}$, one constructs the solution $\Psi_E^c={\hat Q}\Psi_E^c+{\hat P}\Psi_E^c$ in the total function space:
\begin{eqnarray}
  \Psi_E^c=\zeta_E^c+\sum_{i,k}(\Phi_i^A+\omega_i^{(+)}(E))\langle\Phi_i^A|
  (E-{\cal
    H}_{QQ}(E))^{-1}|\Phi_k^A\rangle\langle\Phi_k^A|H_{QP}|\zeta_E^c\rangle 
\end{eqnarray}
${\cal H}_{QQ}(E)$ stands for the energy-dependent OQS Hamiltonian in ${\cal Q}$ subspace:
$  {\cal H}_{QQ}(E)=H_{QQ}+H_{QP}G_P^{(+)}(E)H_{PQ}~,$
which includes the coupling to scattering states and decay channels. In general, this coupling is non-hermitian, and depends both on $E$ and $H_{QP}$. Eigenfunctions ${\tilde \Phi_i^A}$ of the OQS Hamiltonian ${\cal H}_{QQ}$ are linear combinations of SM eigenfunctions $\{\Phi_i^A\}$.

\subsection{The method of anamenses}

Construction of the many-body basis is a key problem in the SM description of OQSs. In Hilbert space, this construction is based on the determination of a complete s.p. basis which consists of  discrete states and a scattering continuum. A consistent formulation of nuclear structure and reactions becomes possible in CSM/SMEC if s.p. resonances are removed from the scattering continuum to be put in the subspace of discrete states  \cite{Bar77}. This regularization procedure for s.p. resonances is associated with an extraction of localized part of s.p. resonances from the scattering continuum which leaves only non-resonant scattering states.  Recently, this problem has been solved by the parameter-free method of anamneses \cite{Fae08}.

For a given pole of the scattering matrix at $k^{res}$, scattering states with 
$\kappa\simeq \Big{\{}{\rm Re}(k^{res})^{2}\Big{\}}^{1/2}$ in the inner region ($r<R$) of a basis generating potential resemble  a bound state wave function. Hence, for a construction of a quasi-bound state embedded in a continuum (QBSEC), it is quite natural to select the scattering state with a (real) energy $e^{res}=\hbar^{2}\kappa^{2}/2\mu$. In the inner region $[0,R]$, where $R$  is yet arbitrary matching radius for inner and outer solutions, the QBSEC  should be proportional to the regular solution of the Schr\H{o}dinger equation with $k=\kappa$. In the external region ($[R,\infty[$), one requires that the QBSEC has the bound state asymptotic corresponding 
to $k=i\kappa$. Since the wave number is fixed, both in the inner $[0,R]$ and outer 
$[R,\infty[$ regions, therefore the continuity of the QBSEC wave function and its first derivative 
at $r=R$ yields a unique solution for $R$, providing an unambiguous determination of the QBSEC. 

Extracting such QBSECs from the scattering continuum allows to determine a new orthogonal subspace of s.p. states, consisting of bound states and QBSECs, and a new subspace of non-resonant scattering states. Such a  s.p. basis is complete and can be directly used for the construction of a complete many-body basis in Hilbert space. It should be stressed that QBSECs constructed in this way yield an {\em anamnesis} (a reminder) of resonances in the space of ${\cal L}^2$-functions, removing all resonant aspects from calculated phase shifts and providing a reliable continuation of weakly-bound s.p. states into a low-energy continuum. This aspect is crucial to achieve a correct description of the near-threshold behavior of many-body states.

\subsection{The Hamiltonian}

%
\begin{figure}[t]\centering
\includegraphics[width=65mm,angle=-90]{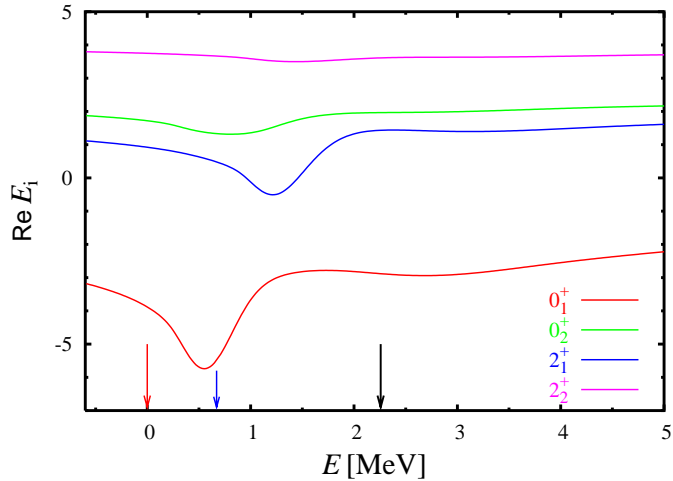}
\caption{Eigenenergies of the effective Hamiltonian are plotted for the two lowest $0^+$ and $2^+$ eigenstates of $^{16}$Ne. Arrows denote positions of the lowest 1p-emission thresholds: 
$\Big{[}^{15}{\rm F}(1/2^+)\otimes{\rm p}(l_j)\Big{]}^{J^{\pi}}$ (red arrow), 
$\Big{[}^{15}{\rm F}(5/2^+)\otimes{\rm p}(l_j)\Big{]}^{J^{\pi}}$ (blue arrow),
$\Big{[}^{15}{\rm F}(1/2^-)\otimes{\rm p}(l_j)\Big{]}^{J^{\pi}}$ (black arrow). For more details, see the description in the text.}
\label{fig22}
\end{figure}
In this work, we study the one-proton (1p) continuum of $^{16}$Ne to look for a possible existence of an EP in a physically relevant region of energies and parameters of the SMEC.  Details of SMEC calculations can be found elsewhere \cite{[Oko03],SMEC,SMEC_2p}. For the effective interaction 
in $H_{QQ}$ we take the ZBM Hamiltonian \cite{ZBM}. The residual couplings between $Q$ and the embedding continuum ($H_{QP}$) is given by: $V_{12}=V_0\delta(r_1-r_2)$ and the strength ($V_0=1100$ MeV$\cdot$fm$^3$) is adjusted to reproduce the experimental width of $5/2_1^+$ state in $^{15}$F. 

The real parts of eigenvalues of ${\cal H}_{QQ}$ are shown in Fig. \ref{fig22} as a function of the total energy of $^{16}$Ne. Zero on the energy scale $E$ is fixed at the first 1p-emission threshold 
$\Big{[}^{15}{\rm F}(1/2^+)\otimes{\rm p}(l_j)\Big{]}^{J^{\pi}}$. Effective Hamiltonian is energy dependent due to the coupling of discrete states of $^{16}$Ne to the embedding 1p continuum
$\Big{[}^{15}{\rm F}(K^{\pi'})\otimes{\rm p}_E(l_{j})\Big{]}^{{J}^{\pi''}}$ which explicitly depends on the total energy of $^{16}$Ne. For $E>0$, ${\cal H}_{QQ}$  is complex-symmetric and its eigenvalues are complex. Opening of successive 1p-emission thresholds (see arrows in Fig. \ref{fig22})
makes a significant change of $0_1^+$, $2_1^+$, $0_2^+$ eigenvalues. Coulomb interaction shifts systematically the near-threshold minima of eigenvalues $E_i(E)$ slightly above the corresponding energies of the 1p-emission thresholds. The main contribution to SM eigenstate $0_1^+$ comes from a coupling to the channel $\Big{[}^{15}{\rm F}(1/2^+)\otimes{\rm p}(1s_{1/2})\Big{]}^{0^+}$. In $2_1^+$ state, the coupling to the channel $\Big{[}^{15}{\rm F}(5/2^+)\otimes{\rm p}(1s_{1/2})\Big{]}^{2^+}$ dominates. In both cases, the dominant continuum coupling is in the $s$-wave. In the excited $0_2^+$ state, the contribution of couplings to the ground state ($1/2_1^+$) and the first excited state ($5/2_1^+$) is about the same.

\section{The exceptional point in the continuum of $^{16}$Ne}

\begin{figure}[t]\centering
\includegraphics[width=60mm,angle=-90]{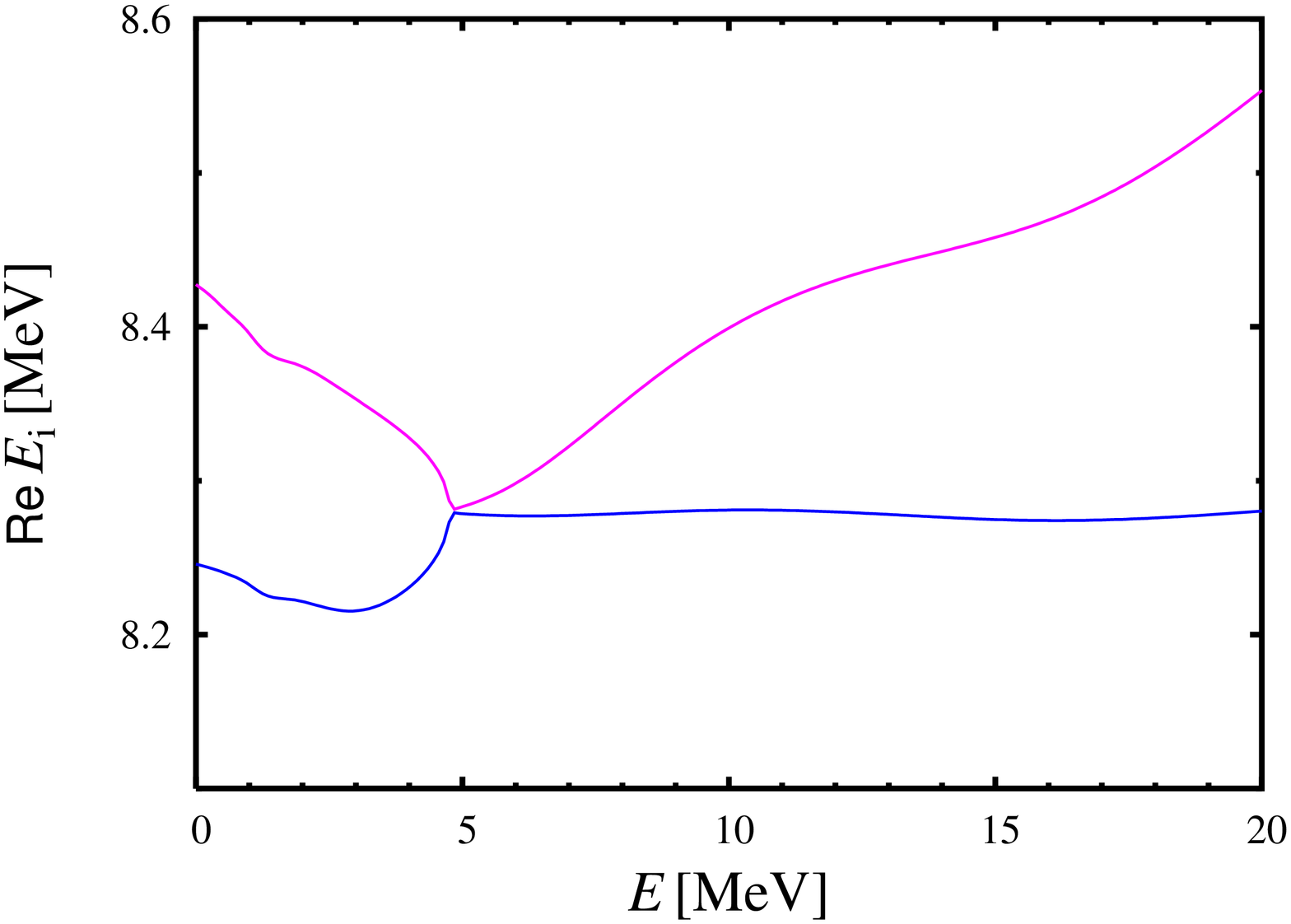}
\includegraphics[width=60mm,angle=-90]{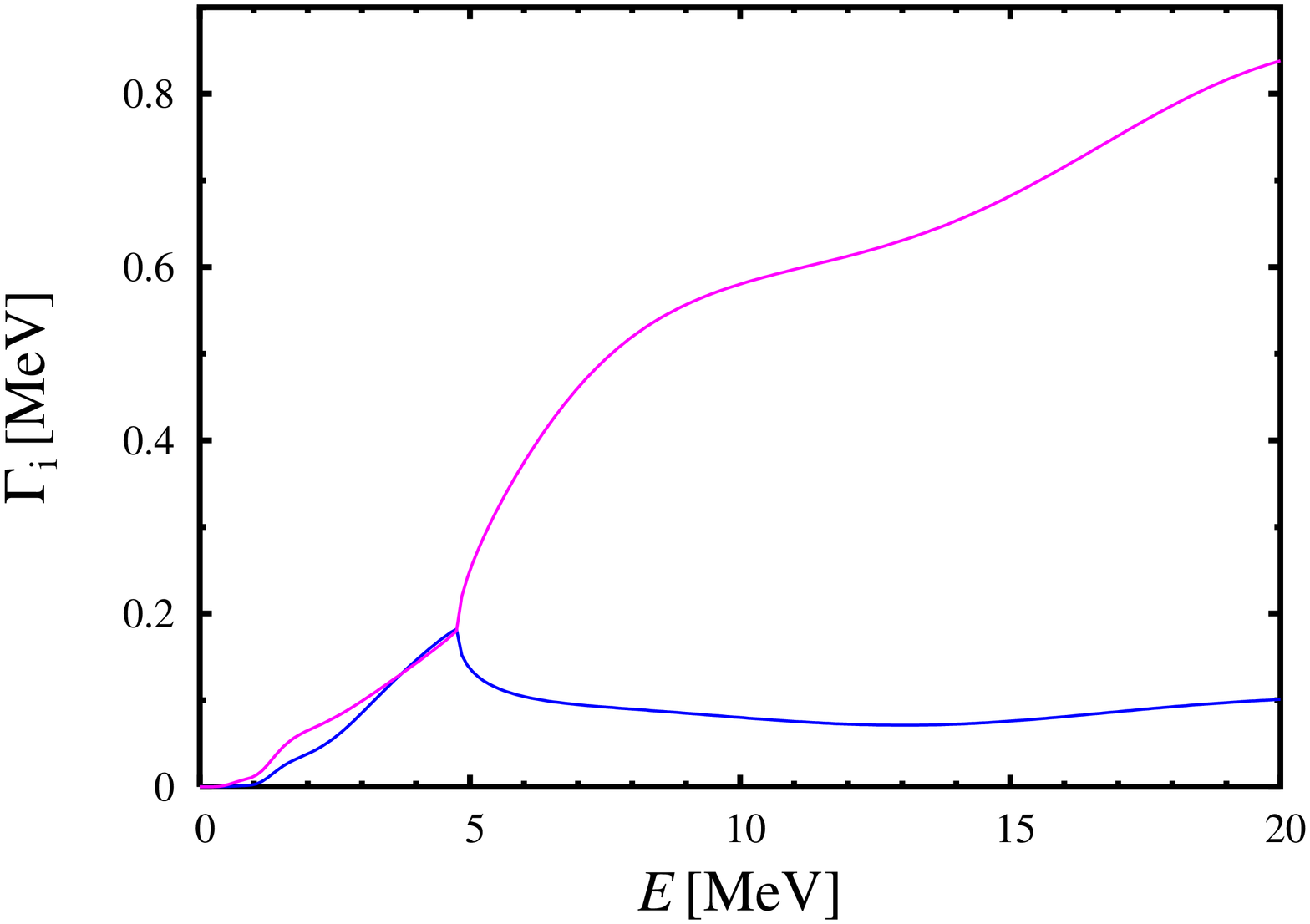}
\caption{Energy and width of $2_3^+$, $2_4^+$ eigenstates of an effective Hamiltonian ${\cal H}_{QQ}$ in $^{16}$Ne which form an exceptional point at a total energy $E=4.765$ MeV. The continuum-coupling strength corresponding to this degeneracy is $V_0=-1357.56$ MeV$\cdot$fm$^3$. For more details, see the description in the text.} 
\label{fig23}
\end{figure}
%
\begin{figure}[t]\centering
\includegraphics[width=60mm,angle=-90]{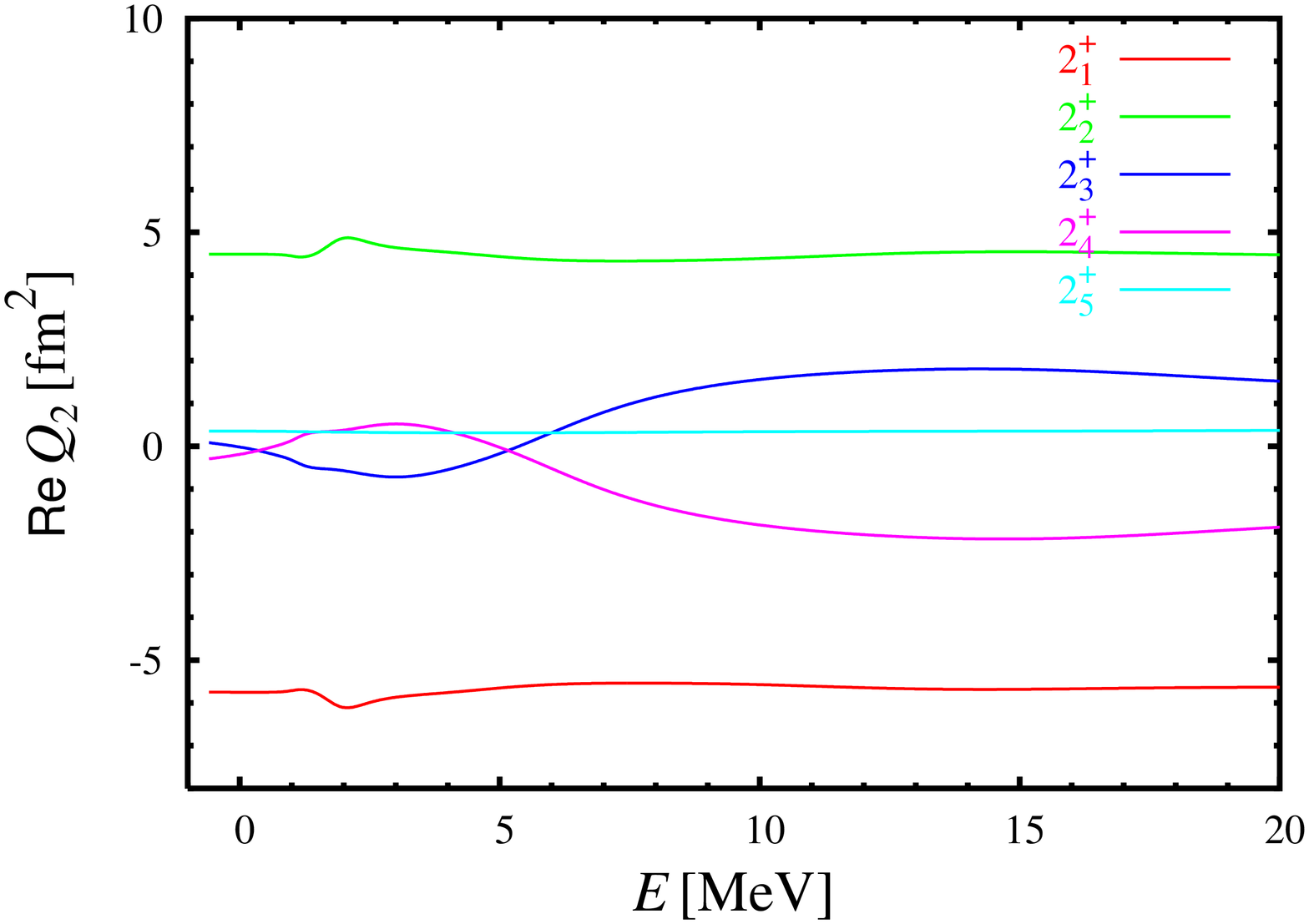}
\includegraphics[width=60mm,angle=-90]{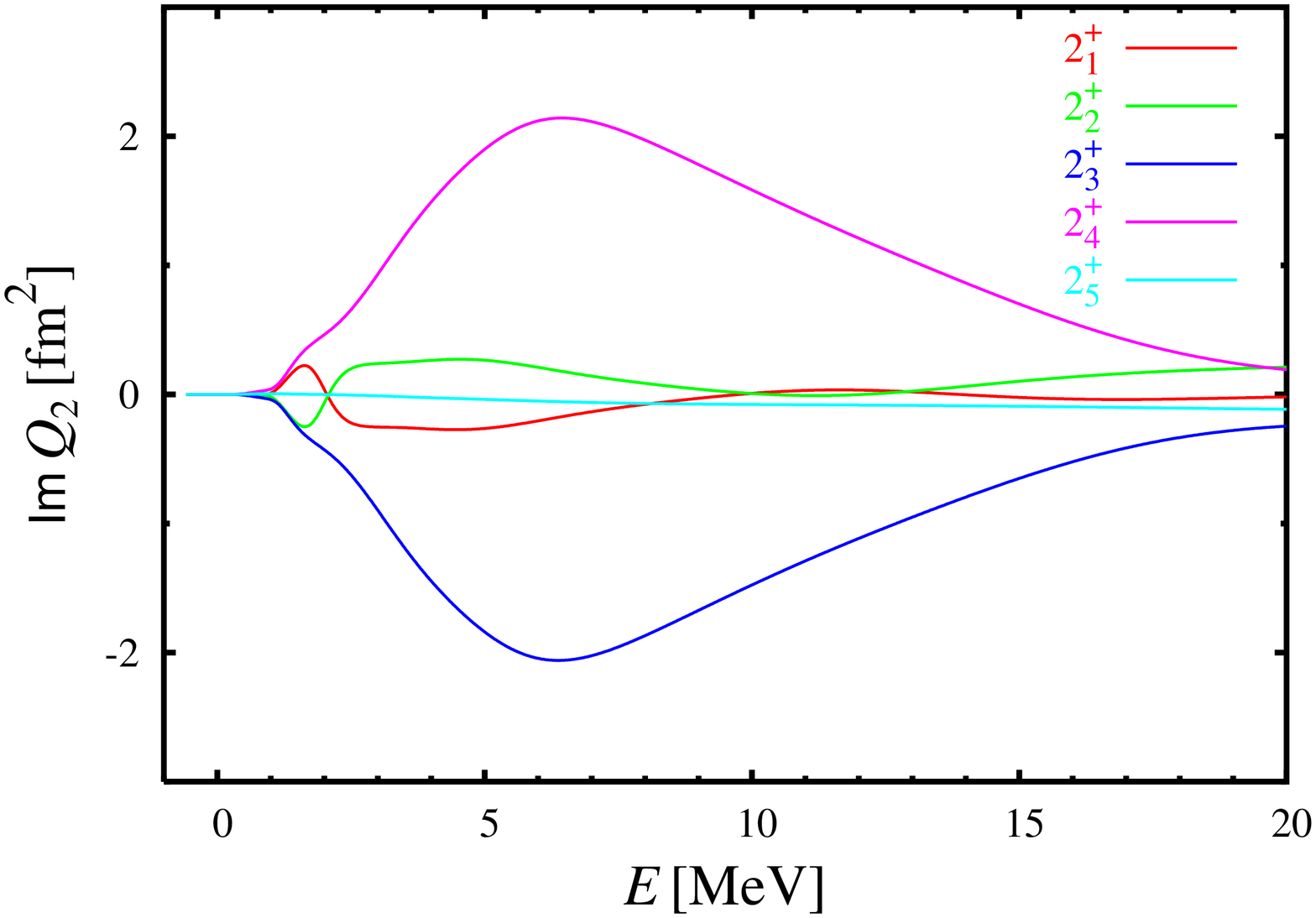}
\caption{Real and imaginary parts of the quadrupole moment for different $2_i^+$ eigenstates of the effective Hamiltonian in $^{16}$Ne. The calculation has been performed for  
$V_0=-1100$ MeV$\cdot$fm$^3$. The index '$i$' enumerates $2^+$ states in ascending order of energy.} 
\label{fig24}
\end{figure}
%
\begin{figure}[t]\centering
\includegraphics[width=52mm,angle=-90]{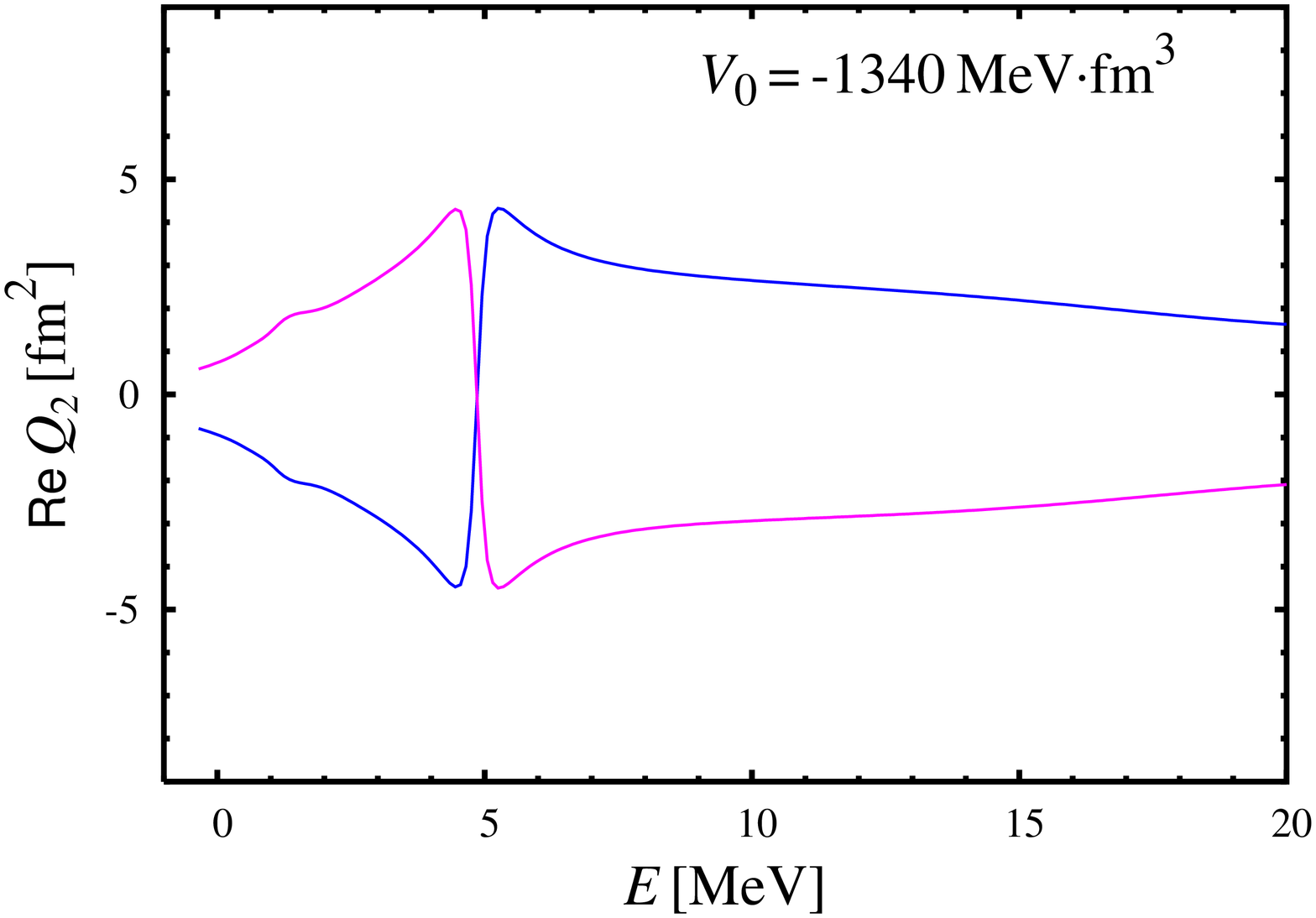}
\includegraphics[width=52mm,angle=-90]{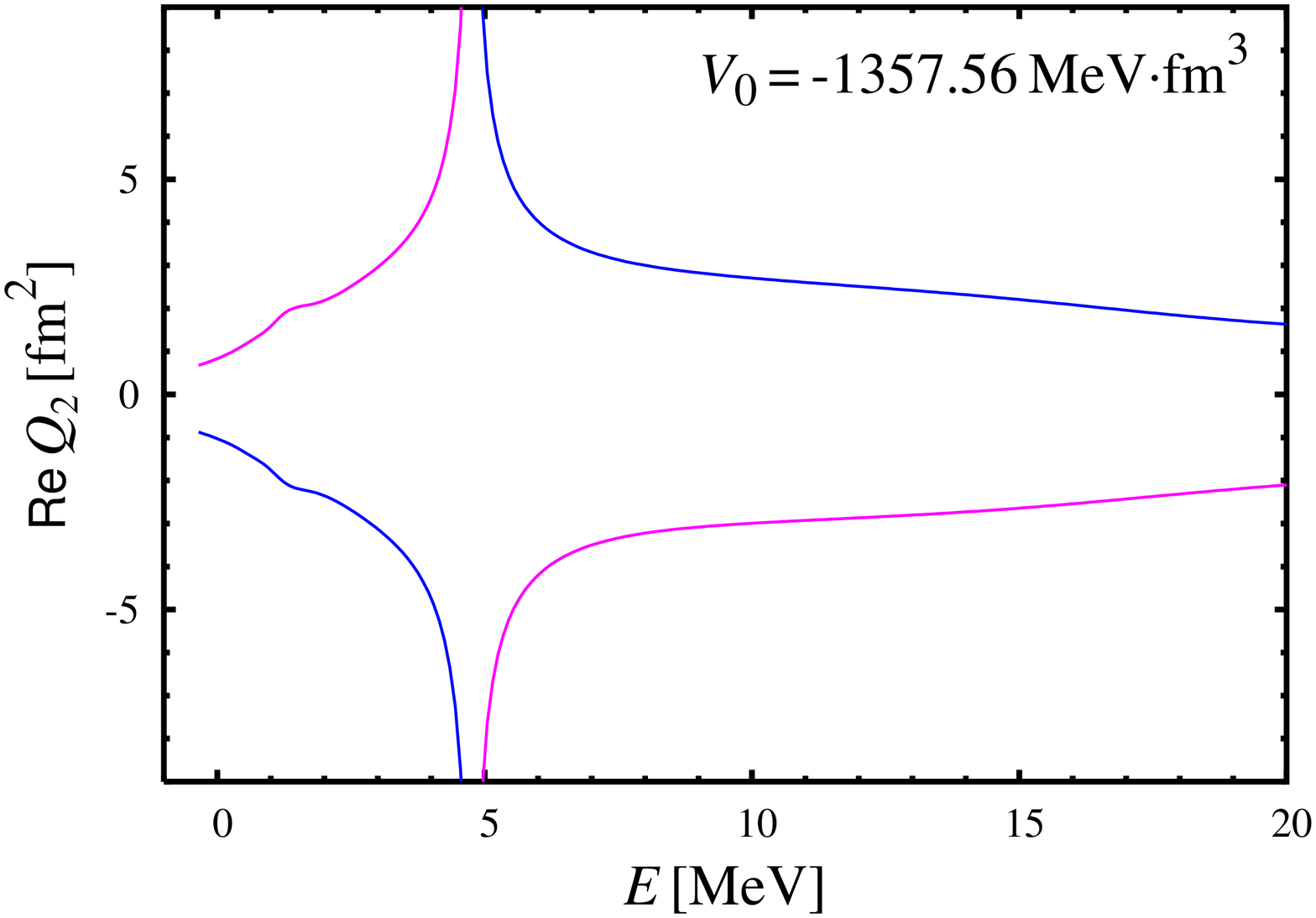}
\includegraphics[width=52mm,angle=-90]{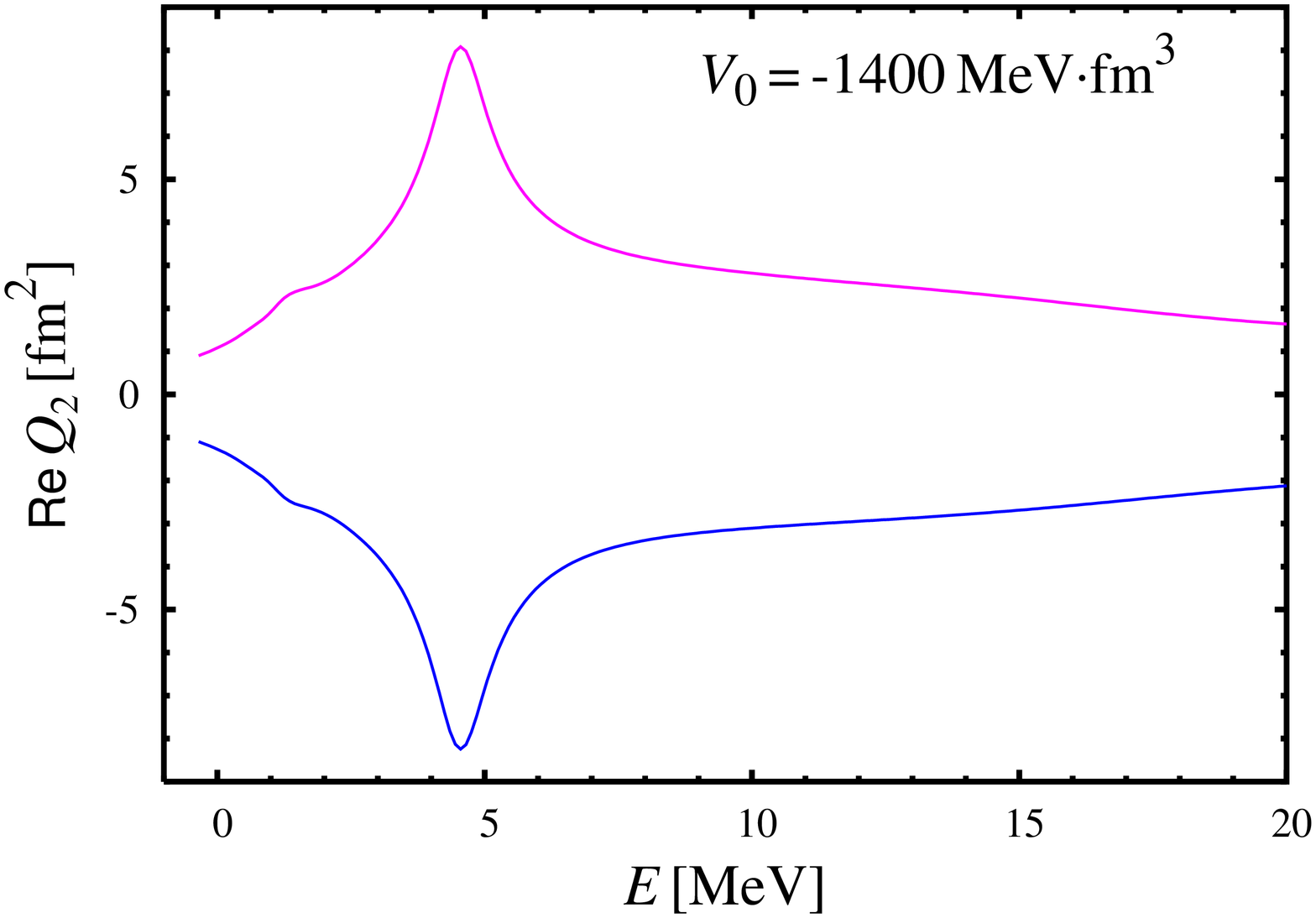}
\caption{Real part of the quadrupole moment for $2_3^+$,  $2_4^+$ eigenstates of an effective Hamiltonian ${\cal H}_{QQ}$ which are involved in the exceptional point at $E=4.765$ MeV
are plotted as a function of $E$ for three values of the continuum-coupling strength: $V_0=-1340, -1357.56$ (the EP at $E=4.765$ MeV), -1400  MeV$\cdot$fm$^3$. For further details, see the discussion in the text.} 
\label{fig25}
\end{figure}
In contrast to diabolic points, which may arise in hermitian systems when two (or more) real parameters are suitably chosen in a Hamiltonian, the appearance of EPs is a generic feature of Hamiltonian systems.  EP appears in the complex $g$ plane of a Hamiltonian $H(g)=H_0+g H_1$, where both $H_0$ and $H_1$ are hermitian and $[H_0,H_1] \neq 0$. The position of degeneracies are indicated by the common roots of equations:
\begin{eqnarray}
{\rm det}\left[H\left(  g\right)  -EI\right]  = 0~;~~~~ \frac{\partial}{\partial E} {\rm det}\left[ H\left(
g\right)  -EI\right] = 0
\label{discr}
\end{eqnarray}
The maximal number of these roots is  $M\leq n(n-1)$, where $n$ is the dimension of the vector space (the number of many-body states of given quantum numbers). Most common degenerate eigenvalues are single-root (EP) and  double-root (diabolic point or level crossing) solutions of Eqs. (\ref{discr}). 

The non-hermiticity of an operator is essential in this context. Whereas a hermitian Hamiltonian can be always diagonalized and eigenvalues, even if degenerate, always correspond to distinct eigenvectors, in non-hermitian Hamiltonian one may find non-trivial Jordan blocks and points in a parameter space of the Hamiltonian where both eigenvalues and eigenvectors coalesce. These EPs have drastic effects on systems behavior, especially concerning geometric phases and adiabatic features. In contrast to DPs, which have been intensely studied in nuclear physics \cite{Ring}, the relevance or even the existence of EPs has never been demonstrated in a realistic nuclear model. Below, we shall present the first study of EPs in the framework of a realistic CSM for unbound states of $^{16}$Ne. 

The EPs are searched here in $E-V_0$  plane, where a variation of the non-hermitian continuum-coupling term can be fully explored. Fig. \ref{fig23} shows energies and widths of 
$2_3^+$, $2_4^+$ resonances as a function of the total energy of $^{16}$Ne for a fixed value of the  strength  $V_0=-1357.56$ MeV$\cdot$fm$^3$ of the $H_{QP}$ coupling term.  $2_3^+$, $2_4^+$ eigenstates and eigenvectors of the effective Hamiltonian coalesce at $E=4.765$ MeV, forming an EP. At this singularity, the dimension $D$ of a vector space decreases ($D=n-1$), and eigenvectors acquire a non-trivial topological phase if the EP is encircled. 

Contrary to the well-known level repulsion mechanism in hermitian system, the entanglement of eigenfunctions is seen far away from the EP and becomes an important source of configuration mixing in the scattering continuum. Fig. \ref{fig24} shows the evolution of real and imaginary parts of the quadrupole moment ${\cal Q}_2$ for five lowest $2_i^+$ states of $^{16}$Ne. In these calculations, we have taken a realistic value of the continuum-coupling strength $V_0=-1100$ MeV$\cdot$fm$^3$, away from the value ($V_0=-1357.56$ MeV$\cdot$fm$^3$) where an EP exists. Imaginary part of  ${\cal Q}_2$ in continuum states, is a measure of the uncertainty of a real part. 

One can notice two distinct phenomena: the threshold effect and the EP effect. Particle-emission threshold is a branch point in OQSs. In Fig. \ref{fig24}, one may notice a strong effect of the coupling to decay channels $\Big{[}^{15}{\rm F}(5/2^+)\otimes{\rm p}(l_j)\Big{]}^{2^+}$
for a pair of eigenstates $2_1^+$, $2_2^+$. The threshold effect is well localized in $E$ and leads to mutually compensating variations in $\langle 2_1^+|{\hat{\cal Q}}_2|2_1^+\rangle$ and 
$\langle 2_2^+|{\hat{\cal Q}}_2|2_2^+\rangle$. The effect of an EP at $E=4.765$ MeV 
($V_0=-1357.56$ MeV$\cdot$fm$^3$) is seen in the eigenstates $2_3^+$, $2_4^+$. The EP effect is weakly localized in energy $E$ and coupling strength $V_0$, and leads to a radical change of 
$2_3^+$, $2_4^+$ eigenfunctions, as manifested by a complicate dependence of 
$\langle 2_3^+|{\hat{\cal Q}}_2|2_3^+\rangle$, 
$\langle 2_4^+|{\hat{\cal Q}}_2|2_4^+\rangle$ on the total energy $E$. 

The intricate dependence of many-body wave functions in the neighborhood of an EP can be seen in Fig. \ref{fig25} on the example of $E$-dependence of a real part of the quadrupole moment 
Re${\cal Q}_2$ in $2_3^+$, $2_4^+$ eigenstates. At a 'critical' value of the continuum-coupling strength 
($V_0=-1357.56$ MeV$\cdot$fm$^3$), at which the EP is found, the quadrupole moments 
$\langle 2_3^+|{\hat{\cal Q}}_2|2_3^+\rangle$, $\langle 2_4^+|{\hat{\cal Q}}_2|2_4^+\rangle$ exhibit a singular behavior, whereas their sum remains finite and smoothly changing around the EP.  This behavior is a manifestation of a genuine non-separability of these two continuum wave functions. It is interesting to notice, that the EP effect remains even if $V_0$ is changed by 5-10\% away from its critical value ($V_0=-1357.56$ MeV$\cdot$fm$^3$), {\em i.e.} the EP effect is rather robust and weakly dependent on a fine-tuning of the Hamiltonian parameters.

\section{Summary and conclusions}

In this work, we have demonstrated that the phenomenon of entanglement of two continuum wave functions by the square-root type of singularity can be found for physically relevant parameters of the nuclear many-body Hamiltonian at low and moderately high excitation energies. This genuine effect in Hamiltonian OQSs is robust and can be seen in a broad range of continuum-coupling strengths around the critical value. Moreover, the entanglement of wave functions remains visible even far away from the exact excitation energy of the EP. These features found in realistic SMEC calculations allow to hope that the indirect consequences of the EP effect can be actually studied in nuclear physics experiments. Further work is needed to determine the most useful reaction/structure observables to study manifestations of the nuclear EP effect.

\section*{Acknowledgements}
We wish to thank J. Dukelsky and W. Nazarewicz for useful discussions.


\begin{thebibliography}{99}
\bibitem{Kle85} P. Kleinw\H{a}chter and I. Rotter, Phys. Rev. C {\bf 32}, 1742 (1985).  
\bibitem{Dro00} S. Dro\.zd\.z {\em et al.}, Phys. Rev. C {\bf 62}, 4313 (2000).
\bibitem{[Oko03]} J. Oko{\l}owicz, M. P{\l}oszajczak, and I. Rotter, Phys. Rep. {\bf 374}, 271 (2003).
\bibitem{Ike68}  K. Ikeda, N. Takigawa, and H. Horiuchi, Prog. Theor. Phys. Suppl. Extra Number, 464 (1968).
\bibitem{Cha06} R. Chatterjee, J. Oko{\l}owicz, and M. P{\l}oszajczak, Nucl. Phys. A {\bf 764}, 528 (2006).
\bibitem{Dob07} J. Dobaczewski {\em et al.}, Prog. Part. Nucl. Phys. {\bf 59}, 432 (2007).
\bibitem{Mic07} N. Michel, W. Nazarewicz, and M. P{\l}oszajczak, Phys. Rev. C {\bf 75}, 031301 (2007).
\bibitem{Oko08} Yan-an Luo {\em et al.}, ArXiv:nucl-th/0211068;\\
J. Oko{\l}owicz, M. P{\l}oszajczak, and Yan-an Luo, Acta Phys. Pol. {\bf  39}, 389 (2008).
\bibitem{Rot00} I. Rotter {\em et al.}, Phys. Rev. E {\bf 62}, 450 (2000). 
\bibitem{SMEC} K.~Bennaceur {\em et al.}, Nucl. Phys. A {\bf 651}, 289 (1999) ; Nucl. Phys. A {\bf 671}, 203 (2000). 
\bibitem{SMEC_2p} J.~Rotureau, J.~Oko{\l}owicz, and M.~P{\l}oszajczak, Phys. Rev. Lett. {\bf 95}, 042503 (2005) ; Nucl. Phys. A {\bf 767}, 13 (2006).
\bibitem{Bet02} R. ~Id Betan {\em et al.}, Phys. Rev. Lett. {\bf 89},  042501 (2002). 
\bibitem{Mic02} N.~Michel {\em et al.}, Phys. Rev. Lett. {\bf 89}, 042502 (2002); Phys. Rev. C {\bf 67}, 054311 (2003); Phys. Rev. C {\bf 70}, 064313 (2004). 
\bibitem{Ber84} {M.V. Berry, Proc. R. Soc. London, Ser. A {\bf 392}, 45 (1984).}
\bibitem{Lau94} {H.-M. Lauber, P. Weidenhammer, and D. Dubbers, Phys. Rev. Lett. {\bf 72}, 1004 (1994);\\
D.E. Manolopoulos, and M.S. Child, Phys. Rev. Lett. {\bf 82}, 2223 (1999); \\
F. Pistolesi, N. Manini, Phys. Rev. Lett. {\bf 85}, 1585 (2000); \\
C. Dembowski {\it et al.}, {\bf 86}, 787 (2001).}
\bibitem{Kat95} {T. Kato, {\em Perturbation Theory for Linear Operators} (Springer Verlag, Berlin, 1995).}
\bibitem{Zir83} {M.R. Zirnbauer, J.J.M. Verbaarschot and H.A. Weidenm\"{u}ller, Nucl. Phys. A {\bf 411}, 161 (1983).}
\bibitem{Hei91} {W.D. Heiss, W.-H. Steeb, J. Math. Phys. {\bf 32}, 3003 (1991).}
\bibitem{Hei91a} W.D. Heiss, A.L.Sanino, Phys. Rev. A{\bf 43}, 4159 (1991).
\bibitem{Cej06} S. Heinze {\it et al.}, Phys. Rev. C {\bf 73}, 014306 (2006).
\bibitem{Duk09} J. Dukelsky, J. Oko{\l}owicz, and M. P{\l}oszajczak, to be published.
\bibitem{Duk04} J. Dukelsky, S. Pittel, and G. Sierra, Rev. Mod. Phys. {\bf 76}, 643 (2004).
\bibitem{Ort05} G. Ortiz {\it et al.}, Nucl. Phys. B {\bf 707}, 421 (2005).
\bibitem{Bar77} H. W. Barz, I. Rotter, J. H{\"o}hn, Nucl. Phys. A {\bf 275}, 111 (1977); {\em ibid.} A {\bf 307}, 285 (1977).
\bibitem{Fae08} J-B. Faes, M. P{\l}oszajczak, Nucl. Phys. A {\bf 800}, 21 (2008).
\bibitem{ZBM} A.P. Zuker, B. Buck, and J.B. McGrory, Phys. Rev. Lett. {\bf 21} (1968) 39.
\bibitem{Ring} R.S. Nikam, P. Ring, Phys. Rev. Lett. {\bf 58}, 980 (1987);\\
R.R. Chasman, P. Ring, Phys. Lett. B {\bf  237}, 313 (1980);\\
J. de Boer, C.H. Dasso, and G. Pollarolo, Z. Phys. A {\bf 335}, 199 (1990);\\
R.S. Nikam, P. Ring, and L.F. Canto, Z. Phys. A {\bf 324}, 241 (1986);\\
L.F. Canto {\it et al.}, Phys. Rev. C {\bf 47}, 2836 (1993);\\
K.G. Helmer {\it et al.}, Phys. Rev. C {\bf 48}, 1879 (1993).

\end{thebibliography}
\end{document}